# A deep convolutional neural network approach to single-particle recognition in cryo-electron microscopy

Yanan Zhu[1], Qi Ouyang[1,2,3], Youdong Mao[1,2,4]*

[1]Center for Quantitative Biology, [2]State Key Laboratory for Artificial Microstructure and Mesoscopic Physics, Institute of Condensed Matter Physics, School of Physics, [3]Peking-Tsinghua Center for Life Sciences, Peking University, Beijing 100871, China. [4]Intel Parallel Computing Center for Structural Biology, Dana-Farber Cancer Institute, Department of Microbiology and Immunobiology, Harvard Medical School, Boston, MA 02115, USA.

*Corresponding author.

Youdong Mao, Ph.D.

450 Brookline Ave, CLS 1010

Boston, MA 02215, USA

Tel: +1 617 632 4358. Fax: +1 617 632 4338.

E-mail address: youdong_mao@dfci.harvard.edu (Y. M.).

**Abstract**

**Background:** Single-particle cryo-electron microscopy (cryo-EM) has become a popular tool for structural determination of biological macromolecular complexes. High-resolution cryo-EM reconstruction often requires hundreds of thousands of single-particle images. Particle extraction from experimental micrographs thus can be laborious and presents a major practical bottleneck in cryo-EM structural determination. Existing computational methods of particle picking often use low-resolution templates as inputs for particle matching, making it possible to cause reference-dependent bias. It is critical to develop a highly efficient template-free method to automatically recognize particle images from cryo-EM micrographs.

**Results:** We developed a deep learning-based algorithmic framework, DeepEM, for single-particle recognition from noisy cryo-EM micrographs, enabling automated particle picking, selection and verification in an integrated fashion. The kernel of DeepEM is built upon a convolutional neural network (CNN) of eight layers, which can be recursively trained to be highly "knowledgeable". Our approach exhibits improved performance and high precision when tested on the standard KLH dataset. Application of DeepEM to several challenging experimental cryo-EM datasets demonstrates its capability in avoiding selection of un-wanted particles and non-particles even when true particles contain fewer features.

**Conclusions**: The DeepEM method derived from a deep CNN allows automated particle extraction from raw cryo-EM micrographs in the absence of templates, which demonstrated improved performance, objectivity and accuracy. Application of this novel approach is expected to free the labor involved in single-particle verification, thus promoting the efficiency of cryo-EM data processing.

**Background**

Single-particle cryo-EM images suffer from heavy background noise and low contrast, due to a limited electron dose used in imaging for reducing radiation damage to biomolecules of interest. Hence, a large number of single-particle images, extracted from cryo-EM micrographs, are required to perform reliable 3D reconstruction of the underlying structure. Particle recognition thus represents the first bottleneck in the practice of cryo-EM structure determination. During the past decades, many computational methods have been proposed for automated particle recognition. Most of these are based on template matching, edge detection, image features or neural networks. The template matching methods [1-7] depend on local cross-correlation that is sensitive to noise. A substantial fraction of false positives may result from false correlation peaks. Similarly, both the edge-based [8-9] and feature-based methods [10-12] suffer from a dramatically reduced performance with lowering the contrast of micrographs. A method based on a three-layer pyramidal-type artificial neural network was developed [13-14]. But there is only one hidden layer in the designed neutral network, which is insufficient to extract rich features in single-particle images. A common problem for these automated particle recognition algorithms lies in that they cannot distinguish "good particles" from "bad" ones, including overlapped particles, local aggregates, background noise fluctuation, ice contamination and carbon areas. Thus, additional steps by unsupervised image classification or by manual selection were necessary to sort out "good particles" after initial automated particle picking. For example, TMaCS uses support vector machine (SVM) to classify the particles initially picked by a template-matching method to remove false positives [15].

Deep learning [16,32] is a type of machine learning techniques that focuses on learning multiple levels of feature representations, which makes sense of multi-dimensional data such as images, sound and text. It is a process of layered feature extraction. In other words, features in greater detail can be extracted with moving up the hidden layer to a deeper level using the multiple non-linear transformations [32]. The area of deep learning consists of deep belief networks (DBNs) [17-18], convolutional neural networks (CNNs) [19,31], recurrent neural networks (RNNs) [20] and any other deep network architecture with more than one hidden layer. CNN is a biologically inspired deep, feed-forward neural network, and has demonstrated outstanding performance on speech recognition and image processing, such as handwriting recognition [21], facical detection [22] and cellular image classification [23]. Its unique advantage lies in that the special structure of shared local weights reduces the complexity of the network [31,33]. Multidimensional images can be directly used as inputs of the network, which avoids the complexity for the feature extraction in the reconstructed data [16,31].

The particle recognition problem in cryo-EM is fundamentally a binary image classification problem based on the features of single-particle images. We devised a novel automated particle recognition approach based on deep CNN learning [31]. It was an eight-layer CNN in our DeepEM model, including an input layer, three convolutional layers, three subsampling layers and an output layer (Fig. 1). In this study, we applied this deep-learning approach to tackle the problem of automated template-free particle recognition, with the goal of detecting "good particle" from cryo-EM micrographs taken at a variety of situations, and with an accuracy improved over existing methods.

**Methods**

**Design of the DeepEM model**

Convolutional neural network is a multilayer neural network with incomplete connections. It contains convolutional layers, subsampling layers and fully connected layers in addition to the input and output layers. The convolutional and subsampling layers produce feature maps through repeated applications of the activation function across sub-regions of the images, which represent low-frequency features extracted from the previous layer (Supplementary Fig. 1).

The convolutional layer is the core building block in a CNN, in which the connections are local, but expand throughout the entire input image. This would ensure that the learnt filters produce the strongest response to a spatially local input. The feature maps from the previous layer is convoluted with a learnable kernel. All convolution operation outputs are then transformed by a nonlinear activation function. We use the sigmoid function as the nonlinear activation function:

$$sigmoid(x) = 1/(1 + e^{-x}) \qquad (1)$$

The convolution operations in the same convolutional layer share the same connectivity weights with the previous layer:

$$X_j^{[l]} = sigmoid(\sum_{i \in M_j} X_i^{[l-1]} * W_{ij}^{[l]} + B^{[l]}), \qquad (2)$$

where $l$ represents the convolutional layer; $W$ represents the shared weights; $M$ represents different feature maps in the previous layer; $j$ represents one of the output feature maps and $B$ represents the bias in the layer and "*" represents the convolution operation.

Subsampling, which is designed to subsample their input data to progressively decrease the spatial size of the representation to reduce the number of parameters and the cost of computation in the network, and hence also to control over-fitting [36], is another important concept of CNNs. We computed subsampling average after each convolutional layer using the following expression, which would reduce the spatial size of the representation,

$$X_{ij}^{[l]} = \frac{1}{MN} \sum_m^M \sum_n^N X_{iM+m, jN+n}^{[l-1]} . \qquad (3)$$

Here $i$ and $j$ represent the position of the output map. $M$ and $N$ represent the subsampling size.

The basic network architecture designed in DeepEM contains three convolutional layers, the first, third, and fifth layers, and three subsampling layers, the second, fourth and sixth layers. The last layer is fully connected with the previous layer, which outputs predictions of the input image by the weight matrix and the activation function (Fig. 1).

**Training of the DeepEM model**

Prior to the application of DeepEM to automated particle recognition, the CNN needs to be trained with a manually assembled dataset sampling both particle images (positive training data) and non-particle images (negative training data) (Examples in Fig. 3a, b). Then the well-trained CNN is used to recognize particles from raw micrographs. We use the error back-propagation method [30] to train the network and produce an output of "1" for the true particle images and "0" for the non-particle images. Data augmentation technique [13,23] has shown certain improvement in accuracy for training the CNNs with a large number of parameters. During our DeepEM training, each training particle image is rotated by 90°, 180° and 270°, to augment the size of data sampling by four times. The weights in the CNN model are initialized with a random number between 0 and 1 and are then updated in the process of training. We use the squared-error loss function [30] as the objective function in our model. For a training dataset with the number of *N*, it is defined by equation (4), in which $t_n$ is the target of the *n*-th training image, and $y_n$ is the value of the output layer in response to the *n*-th input training image.

$$E_N = \frac{1}{2N}\sum_{n=1}^{N}\|t_n - y_n\|^2 \ . \qquad (4)$$

During the process of training, the objective function is minimized using error back-propagation algorithm [30], which performs a gradient-based update:

$$\omega(t+1) = \omega(t) - \frac{\eta}{N}\sum_{k=1}^{N} \varepsilon_n \frac{\partial \varepsilon_n}{\partial \omega} \ . \qquad (5)$$

Here $\varepsilon_n = \|t_n - y_n\|$; $\omega(t)$ and $\omega(t+1)$ represent parameters before and after update in an iteration, respectively; and $\eta$ is the learning rate and is set to 1 in this study.

In reality, the experimental cryo-EM micrographs may contain heterogeneous objects, such as protein impurity, ice contamination, carbon areas, overlapped particles and local aggregates. Moreover, since the molecules in single-particle images assume random orientations, significantly different projection structures of the same macromolecule may coexist in a micrograph. These factors make it difficult to assemble a relatively balanced training dataset first, which includes representative positive and negative particle images. The program is prone to miss some real particles in certain views or recognize some unwanted particles that look similar to the real one. We can optimize the training dataset by adding more representative particle images to the original training dataset after testing on a separate of micrographs that are independent from the ones used for assembling the original training dataset, and then re-train the network following the workflow chart shown in Fig. 2. After sufficient iterations of training, the CNN becomes "knowledgeable" in differentiating positive particles from negative ones.

Since the input particle images size may vary in different datasets, one can set different hyper-parameters for each case, including the number of the feature maps, the kernel size of the convolutional layers and the pooling region size for pooling layers. The details of these hyper-parameters used in this study are shown in Table 1. We designed and implemented the convolutional neural network based on the DeepLearnToolbox [25], which is a Matlab toolbox for deep learning development. All algorithms were implemented in Matlab.

**Particle recognition and selection in the DeepEM model**

When a well-trained CNN is used to recognize particles, a square box of pixels is taken as CNN input. Each input image boxed out of a testing micrograph is rotated incrementally, to generate three additional copies of the input image with rotations of 90°, 180° and 270° relative to the un-rotated one. Each copy is used as a separate input to generate a CNN output. The final expectation value of each input image is taken as the average of its four output values from the un-rotated and rotated copies. The boxed area starts from a corner of the testing micrograph and is raster-scanned across the whole micrograph to generate an array of CNN outputs.

We used two criteria to select particles. First, a threshold score must be defined. The boxed image is identified as a candidate if the CNN output score of the particle is above the threshold score. Those particles whose CNN scores are below the threshold are rejected. We used the F-measure [35], which is a measure on the accuracy of a test that combines both precision and recall for binary classification problems, to determine the threshold score in our approach, which is defined by equation (8):

$$F_\beta = (1 + \beta^2) \cdot \frac{precision \cdot recall}{(\beta^2 \cdot precision + recall)} \ , \qquad (8)$$

where $\beta$ is a coefficient weighting the importance of precision and recall. In our method, we use the $F_2$ score, which weights recall higher than precision. The $F_2$-score reaches its best value at 1 and its worst at 0. We define the cutoff threshold at the highest value of the $F_2$-score. Second, candidate images are further selected based on the standard deviation of the pixel intensities. There are often some carbon areas or contaminants in some micrographs where particles initially detected may not be good choices for downstream single-particle analysis. The pixels in these areas usually have higher or lower standard deviations compared with other areas with clean amorphous ices. We set a narrow range of the pixel standard deviation to remove the candidate particles that are contained in these unwanted areas [5,15] (Supplementary Fig. 2).

**DeepEM algorithm workflow**

Learning process

**Input**: Training dataset

**Output**: Trained CNN parameters (weights and bias)

1. Rotate the input particle image three times, each with a 90° increment;
2. Set the output of the positive data as 1, the output of the negative data as 0;
3. Initialize the hyper-parameters;
4. Randomly initialize the weights and bias in each convolutional layer;
5. Train weights and bias through error back propagation algorithm;
6. Test the trained CNN with an independent testing dataset to measure the learning error;
7. Tune the hyper-parameters or optimize the training dataset by adding more representative positive and negative particles from another set of micrographs, which are independent of those used in step 6, to the training dataset;
8. Repeat steps 4-7 until the CNN output from input testing dataset falls below the defined cutoff of learning error.

Recognition Process

**Input**: Micrographs and trained CNN

**Output**: Box files of selected particles in the EMAN2 format for each micrograph

1. Extract a square of particle size starting from a corner of the input micrograph;
2. Rotate the boxed image three times, each with a 90-degree increment;
3. Use the trained CNN to process four copies of the boxed image, including the un-rotated and rotated copies, and average the resulting output scores of the four images;
4. Repeat steps 1-3 by shifting the square area 20 pixels until the whole micrograph has been raster-scanned;
5. Pick the particle candidates based on their scores that are not only local maximum but also above the threshold score;
6. Select particle images based on their standard deviation;
7. Write the coordinates of the selected particle images to the box file.

**Performance evaluation**

We evaluate the performance of the method based on the precision-recall curve [27], which is one of the popular metrics to evaluate the performance of various particle selection algorithms. Higher precision indicates that an algorithm selects a smaller percent of non-particles. Higher recall means that an algorithm selects a greater percent of the true particle images contained in the micrographs. The precision and recall are defined by equations (6) and (7), respectively, where the precision represents the fraction of true positives (TP) over the total particle images selected (TP+FP) and the recall represents the fraction of true particle images selected over all of the true particle images (TP+FN) contained in the micrographs.

$$\text{Precision} = \frac{TP}{TP+FP} \qquad (6)$$

$$\text{Recall} = \frac{TP}{TP+FN} \qquad (7)$$

The precision-recall curve is generated from the algorithm by varying the threshold score used in the particle recognition procedure. With increasing the threshold, the precision would increase and the recall would decrease accordingly. Thus, the threshold is manifested as a balance between the precision and the recall. For a good performance in a particle selection method, both the precision and the recall are expected to achieve a high value at certain threshold.

**DeepEM training on the KLH dataset**

The KLH dataset was acquired from the National Resource for Automated Molecular Microscopy (nramm.scripps.edu). KLH is ~8 MDa with a size of ~40 nm. It consists of 82 micrographs at 2.2 Å/pixel that were acquired on a Philips CM200 microscope at 120 kV. The size of the micrograph is 2048 by 2048 pixels. There are mainly two types of projection views of the KLH complex, the side view and the top view. We boxed the particle images with a dimension of 272 pixels. 800 particle images were manually selected for a positive training dataset. The same number of randomly selected non-particle images were used as a negative dataset from the first fifty micrographs (Fig. 3a). Each image in the training dataset was rotated at 90° increments to create four images to augment the data. We then had four times positive particle images and negative particle images in the training dataset. We also selected some particle images as a testing dataset containing positive and negative data that were not used in the prior training step. The testing dataset was used to test the intermediately trained CNN model (Fig. 2). The accuracy or error of CNN learning output from the testing dataset was used as a feedback to tune the hyper-parameters, including the number of the feature maps, the kernel size for the convolutional layers and subsampling size for the subsampling layers in the network. For the training process, the intensity of each pixel of an input image was input into a neuron of the input layer; and the desired output was set to 1 for the positive data and to 0 for the negative data. Then the error back-propagation algorithm [30] was used to train the network. Throughout the training-testing processes, we tuned the hyper-parameters and updated the training dataset until the accuracy of the CNN learning reaches a satisfying level. The acceptable value was often set as ~95% at the threshold of 0.5 (Fig. 2).

**Application to experimental cryo-EM data**

The original sizes of the micrographs for inflammasome, 19S regulatory particle and 26S proteasome are 7420 by 7676, 3710 by 3838 and 7420 by 7676 pixels, respectively. The pixel sizes the inflammasome, 19S regulatory particle and proteasome holoenzyme are 0.86, 0.98 and 0.86 Å/pixel, respectively. For the inflammasome and 26S proteasome, the micrographs were binned for 4 times. Therefore, the pixel sizes we used for inflammasomes and proteasome holoenzyme are 3.44 Å/pixel, respectively. For the 19S regulatory particle, the micrographs were binned for 2 times, resulting in the pixel size of 1.96 Å/pixel. Thus, the resulting sizes of the micrographs used in our tests are all 1855 by 1919 pixels; the dimension of the particle images of inflammasome, 19S and 26S complexes are 112, 160 and 150 pixels, respectively. These experimental cryo-EM datasets were acquired from an FEI Tecnai Arctica microscope at 200 kV equipped with a Gatan K2 Summit direct electron detector. We applied the DeepEM to these three challenging cryo-EM datasets. The hyper-parameters we tuned for these dataset are shown in Table 1. Different from the training for the KLH dataset, we added the false positive and true positive data, which were manually verified on a separate set of micrographs independent of the testing dataset used for tuning the hyper-parameters, to optimize the training dataset and train the network again for the low-contrast datasets (Supplementary Fig. 3).

**Results**

**Experiments on the KLH dataset**

We first tested our DeepEM algorithm on the Keyhole Limpet Hemocyanin (KLH) [26] dataset that was previously used as a standard testing dataset to benchmark various particle selection methods [2-3,5,7,10-12,15]. For the KLH dataset, the recall and the precision can both reach ~90% at the same time in the precision-recall curve (Fig. 3f) plotted against a manually selected set of particle images from 32 micrographs, which did not include any particle images used in the training dataset. Our approach achieves a higher precision over all of the particle images selected, whereas the recall is still kept at a high value, indicating that fewer false-negative particle images are missed in the micrographs. In a typical KLH micrograph (Fig. 3c), all true particle images are automatically recognized by our method with a threshold of 0.84, which is determined by the $F_2$-score (see Method and equation 8) (Fig. 3e). Comparison of the precision-recall curves between DeepEM with those from RELION and TMACS suggests that DeepEM outperforms those template-matching based methods (Supplementary Fig. 4).

To understand the impact of the training particle number on the algorithmic performance, we varied the particle number in the KLH training dataset from 100 to 1200 and plotted the corresponding precision-recall curves (Fig. 4). In each testing case, the number of positive particles was maintained equal to that of the negative particles. Although there was clear improvement in the precision-call curve when the training particle number was increased from 100 to 400, there was little improvement with further increasing the training data size. The best result was obtained on the training with 800 positive particle images.

**Experiments on challenging cryo-EM datasets**

We also applied our method to several challenging cryo-EM datasets collected with a direct electron detector. Several datasets were examined, including the 19S regulatory particle, 26S proteasome and NLRC4/NAIP2 inflammasome [34]. Figure 3d shows a typical micrograph of 19S, in which DeepEM selects nearly all true particle images contained in the micrograph. Meanwhile, it avoids selecting non-particles from areas of aggregates and carbon film. The precision-recall curve resulting from the test on the 19S dataset is shown in Fig. 3f. The precision and recall both reach ~80% at the same time. The picked particles are approximately well centered as compared to the manually boxed ones. To further verify that the particle images selected are correct, we performed unsupervised 2D classification. The resulting reference-free class averages from about 100 micrographs are consistent with different views of the protein samples (Supplementary Fig. 5).

Two difficult cases from the inflammasome dataset were examined. Figure 4a shows a micrograph with a high particle density and containing excessively overlapped particles and ice contaminants. Most methods based on template matching are incapable of avoiding particle picking from overlapped particle and ice contaminants in this case. Figure 4b presents another difficult situation, in which the side views of inflammasome are of a lower SNR, lack in low-frequency features, and are dispersed with a very low spatial density. In both cases, DeepEM can still perform quite well in particle recognition, while avoiding selection of overlapped particles and non-particles. Further tests on similar cases from other protein samples suggest that this is well reproducible (Supplementary Fig. 6). Importantly, DeepEM has been successfully applied to determine the structures of the human 26S proteasome [37].

**Computational efficiency**

The DeepEM was tested on a Macintosh with 3.3 GHz Intel Core i5 and 32 GB memory, installed with Matlab 2014b. When the size of particle image increases, the parameter space will increase substantially so that it costs more computational time for each micrograph. We usually binned the original micrographs 2 or 4 times to reduce the size of the particle image. For the KLH dataset, it took about 7300 seconds per micrograph with the micrograph size 2048 by 2048 pixels and particle image size 272 by 272 pixels. For 19S regulatory particle, inflammsome and 26S proteasome datasets, it took about 790, 560, and 1160 seconds per micrograph with the binned micrograph size of 1855 by 1919 pixels and the particle image sizes of 112 by 112, 160 by 160, 150 by 150 pixels, respectively. To speed up, multiple instances of the code were run in parallel.

**Discussion**

Based on the principle of deep CNN, we have developed the DeepEM algorithm for single-particle recognition in cryo-EM. The method allows automated particle extraction from raw cryo-EM micrographs, thus improving the efficiency of cryo-EM data processing. In our current scheme, a new dataset containing particles of significantly different features may render the previously trained hyper-parameters suboptimal. Finding a set of fine-tuned hyper-parameters leading to optimized learning results on new datasets demands additional user intervention in CNN training. In the above-described examples, we screened several combinations of the hyper-parameters to empirically pinpoint an optimal setting. This procedure may be not efficient and can be laborious in certain cases. An automated method for systemically tuning hyper-parameters could be developed in the future to address this issue. From the experiments on the inflammasome dataset, we notice that DeepEM is more effective for feature-rich dataset. It exhibits a reduced performance when tested on the side views as compared to the top views of the inflammasome dataset (Fig. 4c), in that the side views exhibit significantly less low-frequency features than the top views. In addition, we also examined the noise tolerance of the model with simulated datasets. When the SNR was decreased to 0.01, the DeepEM can still recognize particle images well after proper training (Fig. 6).

Our DeepEM algorithm framework exhibits several advantages. First, with sufficient training, DeepEM can select true particles without picking non-particles, in a single, integrative step of particle recognition. It performs as well as a human worker. Similar performance was only made possible previously through combining several steps in automated particle picking, unsupervised classification and manual curation. Second, DeepEM features the traits representative of other artificial intelligent (AI) or machine learning systems. The more it is trained or learned, the better it performs. We found that with iteratively updating or optimizing the training dataset, the performance in particle recognition by DeepEM can be further improved. However, this was not possible for conventional particle-recognition algorithms developed so far, whose performance was intricately bound by their mathematics and control parameters.

**Conclusion**

DeepEM derived from deep CNN has proved to be a very useful tool for particle extraction from noisy micrographs in the absence of template. This approach gives rise to improved "precision-recall" performance in particle recognition, as well as demonstrates increased tolerance to much lower SNRs in the micrographs, in contrast to those template-matching based methods. Therefore, it enables automated particle picking, selection and verification in an integrated fashion at a level of a human worker. We expect that this development broadens applications of modern AI technology in expediting cryo-EM structure determination. Related AI technology may be developed in the near future to address key challenges in this area, such as deep classification of highly heterogeneous cryo-EM datasets.

**Abbreviations**

CNN: Convolutional neural network

Cryo-EM: Cryo-electron microscopy

SNR: Signal-to-noise ratio

AI: Artificial intelligence

KLH: Keyhole Limpet Hemocyanin

# Declarations

## Acknowledgements

The authors thank H. Liu, Y. Xu, M. Lin, D. Yu, Y. Wang, J. Wu and S. Chen for helpful discussion and thank S. Zhang for the feedback on the noise tolerance of the DeepEM.

## Funding

The cryo-EM experiments were performed in part at the Center for Nanoscale Systems at Harvard University, a member of the National Nanotechnology Coordinated Infrastructure Network (NNCI), which is supported by the National Science Foundation under NSF award no. 1541959. This work was funded by a grant of the Thousand Talents Plan of China (Y.M.), by grants from National Natural Science Foundation of China No. 11434001 and No. 91530321 (Y.M., Q.O.), by the Intel Parallel Computing Center program (Y.M.).

## Availability of data and material

Our software implementation in Matlab is freely available at http://ipccsb.dfci.harvard.edu/deepem. The experimental micrographs data are freely available at Electron Microscopy Pilot Image Archive (https://www.ebi.ac.uk/pdbe/emdb/empiar/) under accession code: EMPIAR-10063 and EMPIAR-10072.

## Authors' contributions

Conceived and designed the experiments: YZ QO YM. Performed the experiments: YZ. Analyzed the data: YZ YM. Contributed reagents/materials/analysis tools: QO YM. Wrote the paper: ZY YM. All authors have read and approved the final manuscript.

## Competing interests

The authors declare that they have no competing interests.

## Consent for publication

Not applicable.

## Ethics approval and consent to participate

Not applicable.

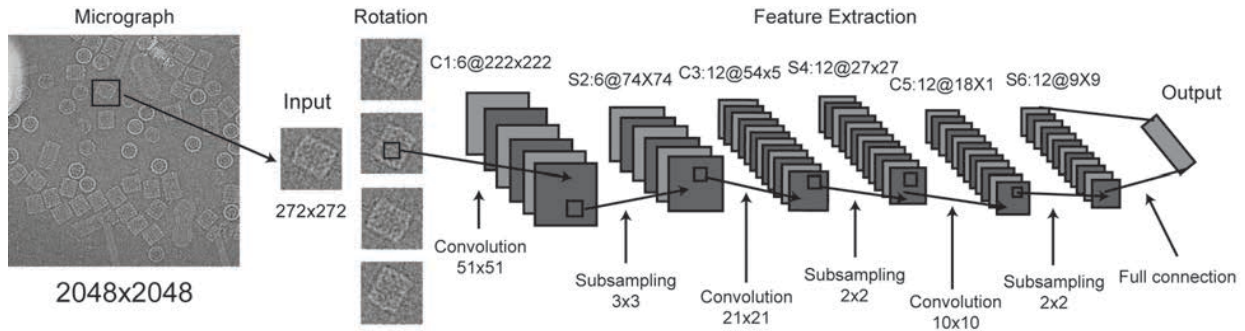

**Figure 1**. **The architecture of the convolutional neural network designed in DeepEM**. The convolutional layer and the subsampling layer are abbreviated as C and S, respectively. C1:6@222×222 means that it is a convolutional layer and is the first layer of the network. This layer is comprised of six feature maps, each of which has a size of 222 × 222. The symbols and number above the feature maps of other layers have the similar meaning.

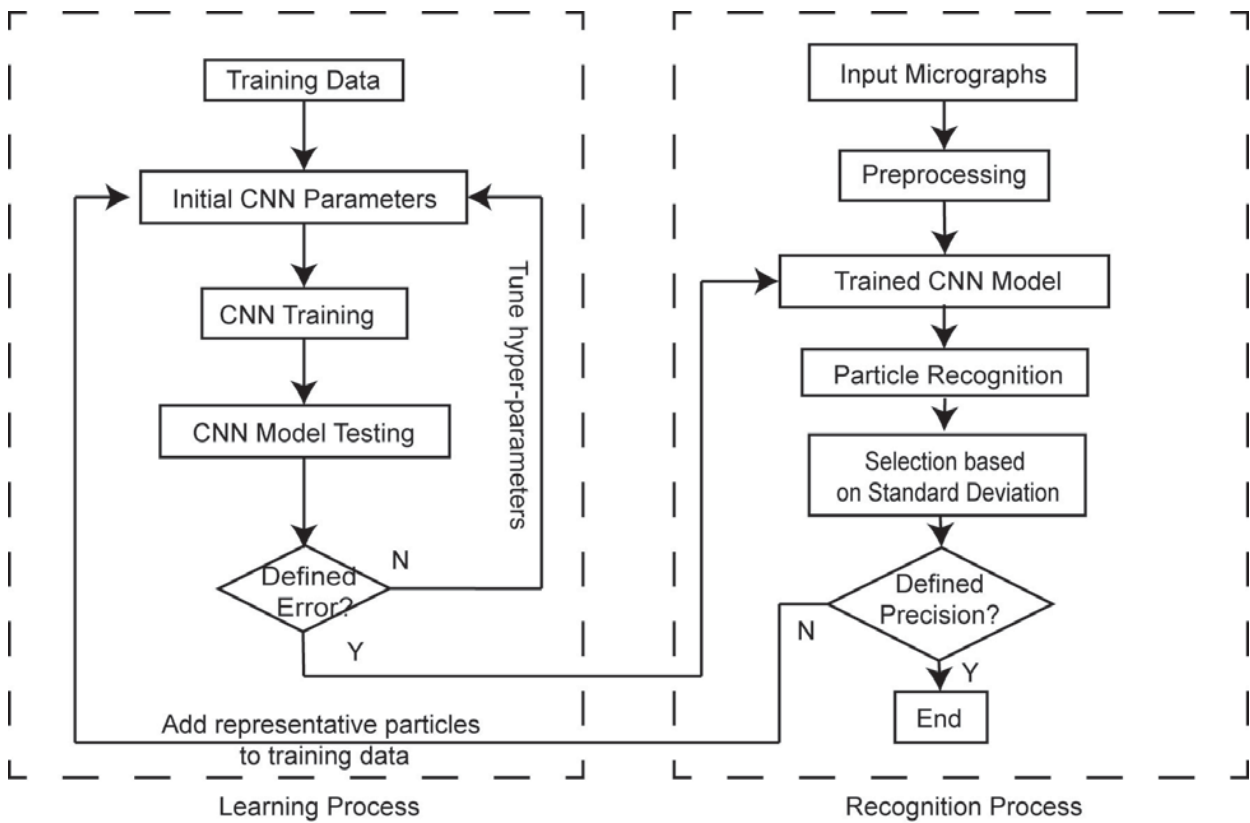

**Figure 2. The workflow chart of the DeepEM.** Left dashed box presents the learning process; and right dashed box presents the recognition process.

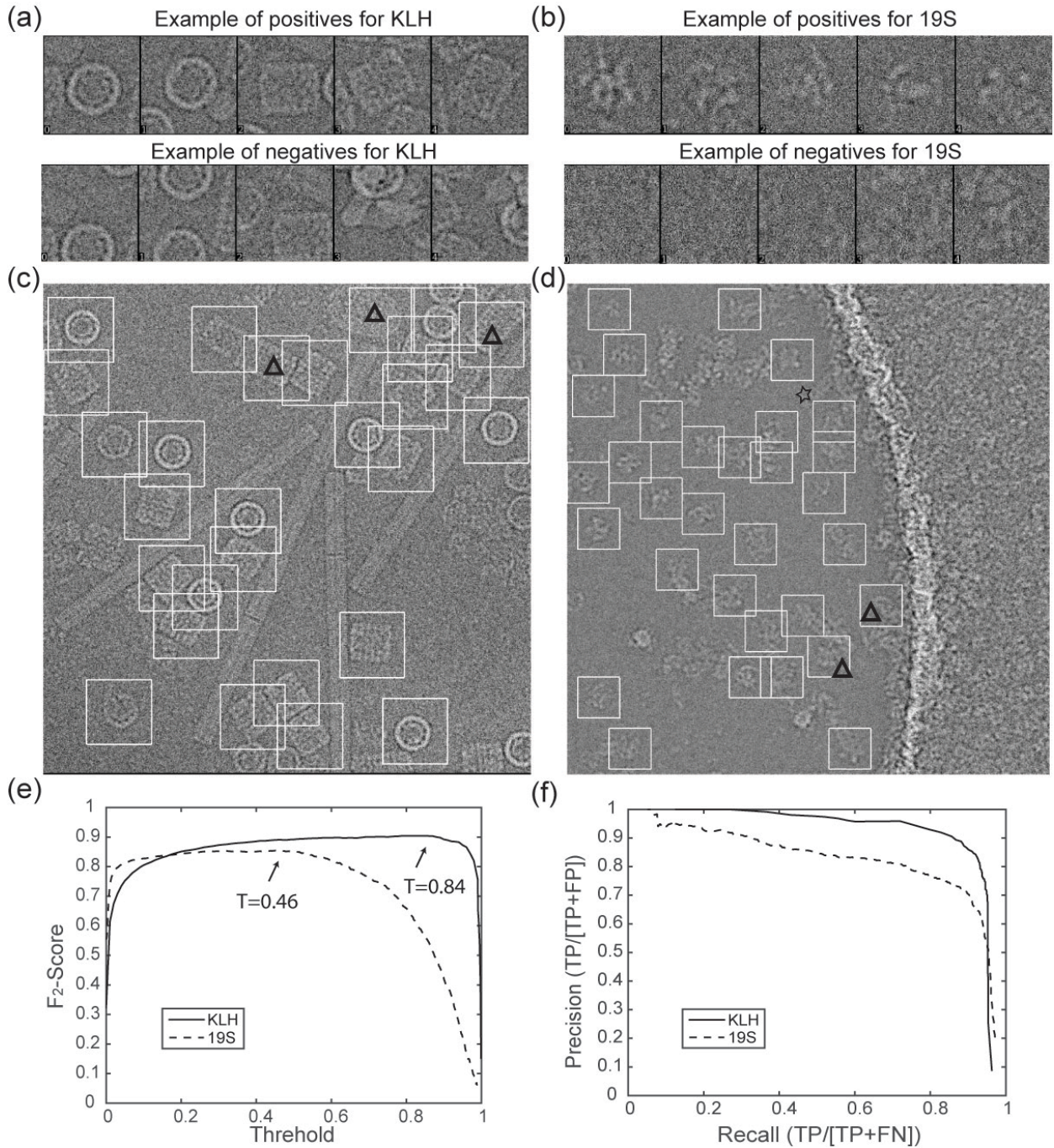

**Figure 3**. **The DeepEM results of KLH dataset and 19S regulatory particles.** (a) and (b) The example of positive and negative particle images selected previous respectively. (c) and (d) Typical micrographs from KLH and 19S dataset, respectively. The white square boxes indicate the positive particle images we selected with the DeepEM program; and boxes with a triangle in them indicate that the false-positive particle images picked. (e) The F$_2$-score curves provide

different threshold for particle recognition in the KLH and 19S datasets, the arrows indicate the peak in each curve, where the cutoff threshold value is defined. (f) The precision-recall curves plotted against manually selected list of particle images.

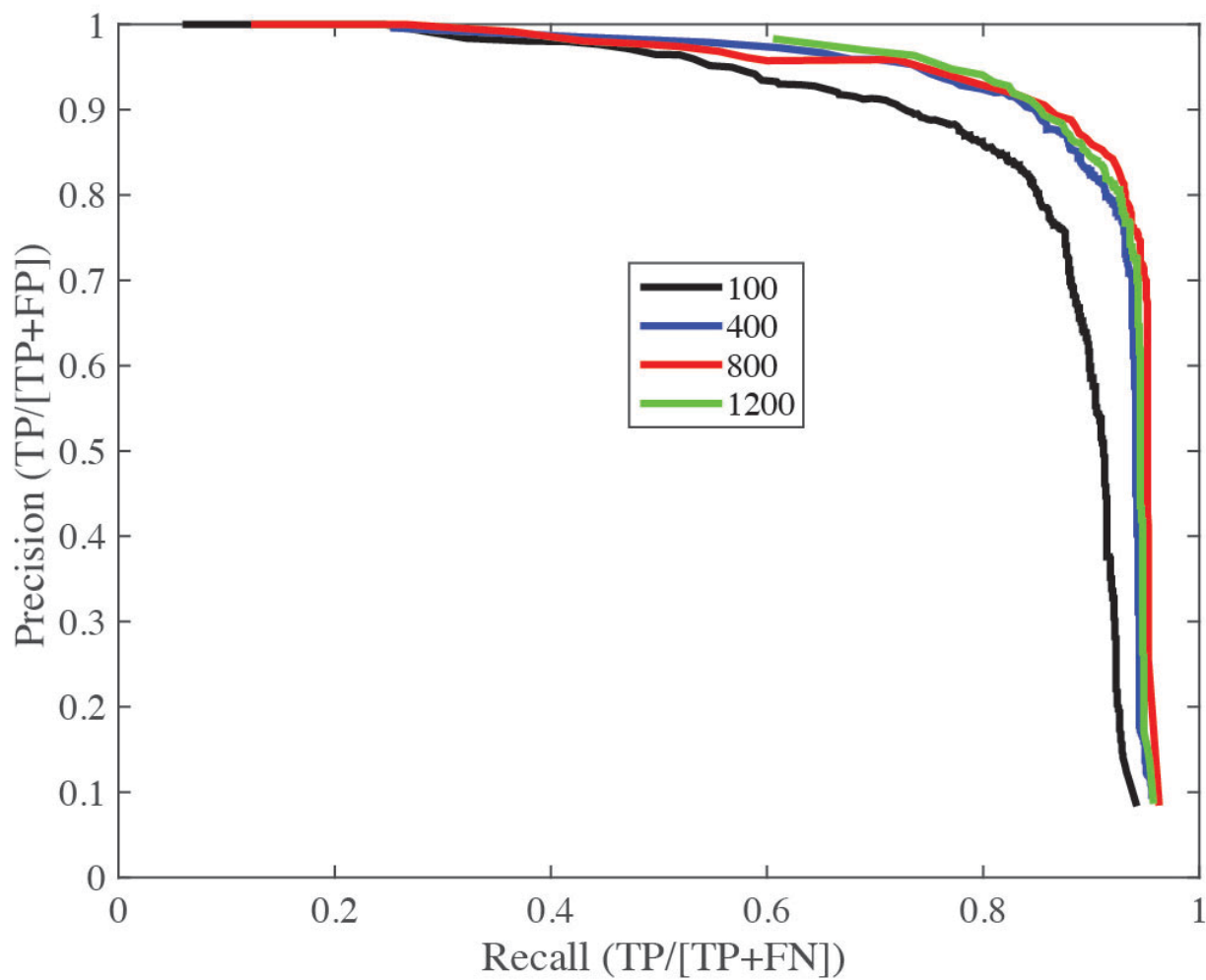

**Figure 4. The impact of the training image number on the precision-recall curve.** The black, blue, red and green curves were obtained with the training datasets including positive image numbers (or negative image numbers) of 100, 400, 800 and 1200, respectively.

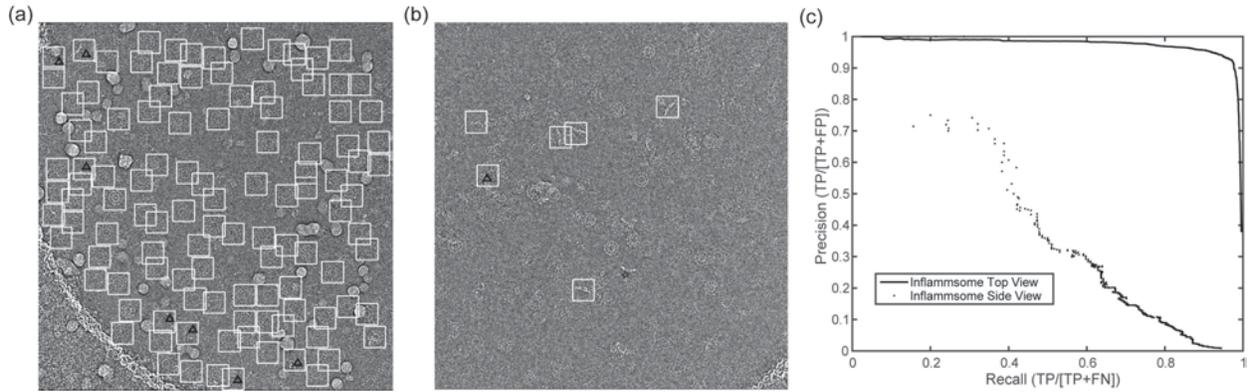

**Figure 5**. **Two challenging examples of automated particle recognition.** (a) A typical example of high-density top view of the inflammasome complex. Considerable ice contaminants and overlapped particles coexist in the micrograph. (b) A typical example of side view showing both few features and a low density of objects. The white square boxes indicate the positive particle images selected by DeepEM. Boxes with a triangle inside indicate that the false-positive particle images picked. Boxes with a star inside indicate the particle images left. (c) The precision-recall curves corresponding to the cases shown in (a) and (b).

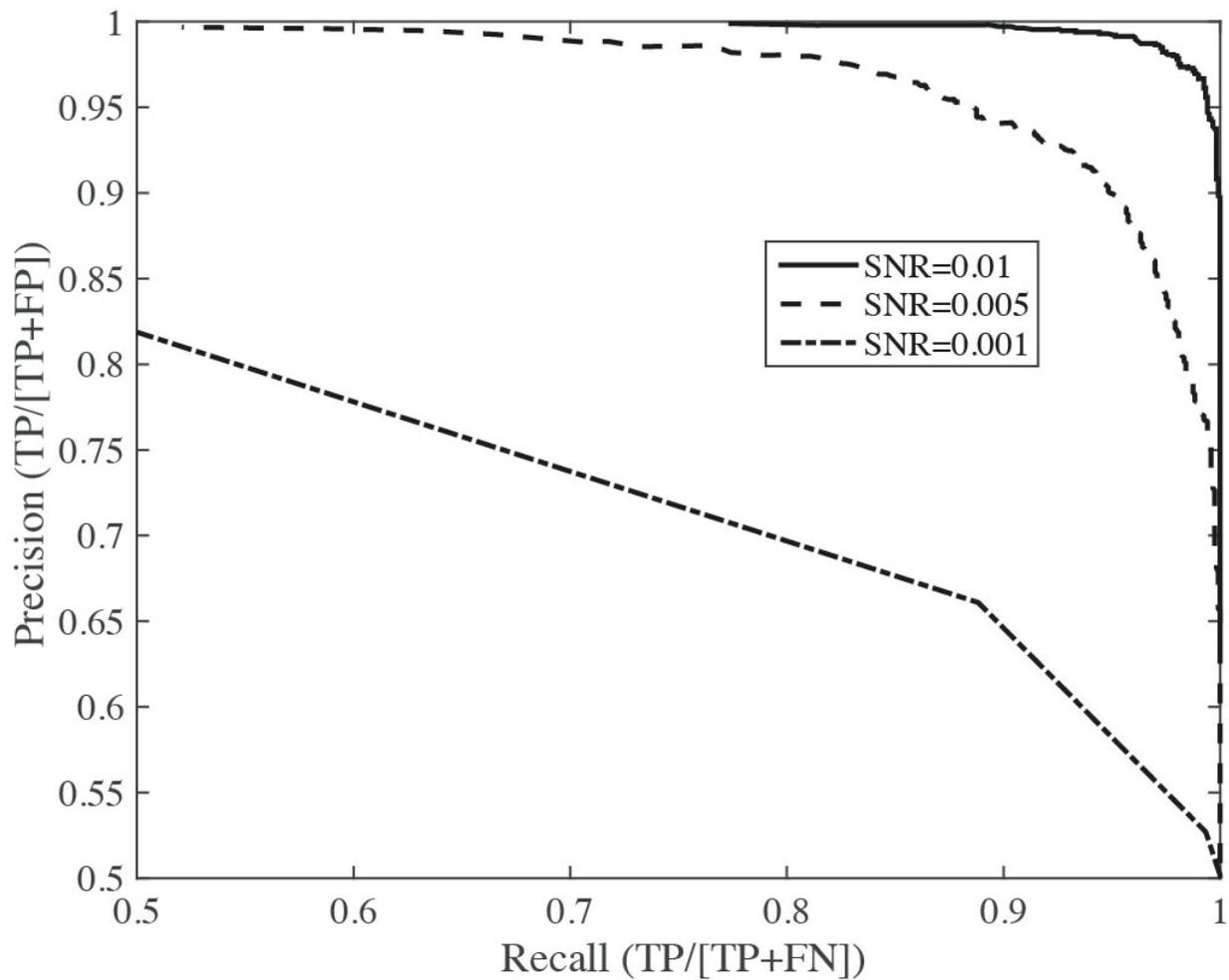

**Figure 6. The effect of the SNR on the precision-recall curves.** Three synthetic datasets were generated through simulation of micrographs containing single-particle images with SNR of 0.01, 0.005 and 0.001. For each case, the CNN was first trained on the synthetic dataset of a given SNR and then used to examine the precision-recall relationship using another synthetic dataset of the same SNR. All synthetic datasets used the 70S ribosome as the single-particle model.

**Table 1. Hyper-parameters used in different datasets**

| Dataset | Particle Size | Corresponding Layer in DeepEM | | | | | |
|---|---|---|---|---|---|---|---|
| | | C1 | S2 | C3 | S4 | C5 | S6 |
| KLH | 272X272 | 6@222X222 | 6@74X74 | 12@54X54 | 12@27X27 | 12@18X18 | 12@9X9 |
| 19S | 160X160 | 6@141X141 | 6@47X47 | 12@38X38 | 12@19X19 | 12@16X16 | 12@8X8 |
| 26S | 150X150 | 6@120X120 | 6@60X60 | 12@46X46 | 12@23X23 | 12@14X14 | 12@7X7 |
| Inflammasome | 112X112 | 6@98X98 | 6@49X49 | 12@40X40 | 12@20X20 | 12@14X14 | 12@7X7 |